\documentstyle[prd,aps]{revtex}

\newcommand{\be}{\begin{equation}}
\newcommand{\ee}{\end{equation}}
\def\bq{\begin{eqnarray}}
\def\eq{\end{eqnarray}}

\def\th{\theta}
\def\Th{\Theta}
\def\de{\delta}

\def\G{\Gamma}

\begin{document}

\title{Non-conformally flat bulk spacetime and the $3$-brane world}

\author{Parampreet Singh$^{a}$ and Naresh Dadhich$^{b}$}
\address{Inter-University Centre for Astronomy and Astrophysics,
Post Bag 4, Ganeshkhind, Pune~411~007, India.}
\maketitle
\begin{abstract}
We show that for non-conformally flat bulk spacetime, there exist no
bound modes for zero mass graviton on the $3$-brane. The brane world model is 
therefore unstable for the bulk spacetime being different from the conformally flat anti - de Sitter space. \\

\noindent
PACS: 04.50.+h, 11.10.Kk
\end{abstract}

\vspace{0.5cm}

 Inspired by the requirement of dimensions more than four for the
 string theory, the study of gravity in higher dimensions has been in
 vogue for quite a while. Generally the extra dimensions are taken to
 be compact and could be probed only at the Planck energy. In view of
 the recent work, it is no longer necessary to have the extra
 dimensions compact, they could be large with matter fields being
 confined to a $3$-brane in $1+3+d$ dimensional bulk [1,2] and gravity
 propagating in the extra dimensions. The brane world model envisages that 
the standard model matter fields live on the $3$-brane while gravity 
propagates in the bulk which is the $Z_2$ symmetric Einstein space. In particular, the most talked of Randall-Sundrum (RS) model [3] has $5$-dimensional anti - de Sitter space as the bulk with brane being Minkowskian flat. It is then shown
 that gravity can, even with infinite fifth dimension, be localized on
 a single $3$-brane.

 In the RS model, the confinement of zero mass graviton has been obtained by the exponential conformal factor in the non-factorizable bulk metric given
by 
\be ds^2 = dy^2 + e^{-2k|y|}\left(-dt^2 + d{\bf x}^2\right) 
\ee 
This is the anti - de Sitter space in the $5$-dimensional bulk for $y\neq 0$
with $\Lambda = -6k^2$. The brane is sitting at the fixed point $y=0$
with $y<0$ being identified by $y>0$ reflecting the $Z_2$
symmetry. The perturbations of the metric do give on the brane the
Newtonian $1/r$ potential with the correction term going as
$1/r^3$. Thus the conventional gravity is recovered on the brane with
the first order correction which could be probed only at sub-millimeter
level [4].

 Recently there have been many works questioning the stability of the brane world model in various contexts [5]. In particular it has been argued
 that there has to be maintained a very delicate balance between the
 $\Lambda$ and the tension in the brane [6]. In the above metric, the
 brane is flat while the bulk is anti - de Sitter. These are undoubtedly
 very special spaces describing very specific idealized situations and
 hence could by no means be considered generic. Recently, it has also
 been shown that the bound state of zero mass graviton is normalizable
 only when the brane spacetime could at best be Ricci flat for spherically symmetric spacetimes. That is,
 the brane space cannot be different from the Schwarzschild solution, not even a de Sitter space [7]. Thus localizability of gravity on the brane is
 accomplished under very special circumstances.

 In this note, we wish to examine what happens if we take the bulk to be a non conformally flat Einstein space with of course negative $\Lambda$. That is it has non-zero Weyl curvature which would induce a trace free stress tensor on the brane. The question is, would there exist a bound state for the zero mass graviton in this case? It turns out that it doesn't. That is what we show in the following.

 It may be noted that one of the most attractive features of the brane world 
model was to have free gravitational field in the bulk reflected on
the brane through the projection of the Weyl curvature as trace free
matter field on the brane. This is how the bulk back reacts to produce ``tidal'' charge on the brane which gets however washed out when the bulk is anti - de Sitter because it has vanishing Weyl curvature (conformally flat). It is therefore pertinent to consider
the solution of the the equation $G_{ab} = -\Lambda g_{ab}$ for the
bulk spacetime with non-zero Weyl curvature. There exists such a
solution obtained due to Nariai [8] for the 4-dimensional Einstein space,
which is conformally non flat. We extend this solution to the
5-dimensional bulk spacetime and it reads as follows: 
\be
 d s^2 = d y^2 + e^{- 2 k |y|} \left( - d t^2 + d r^2 \right) + \frac{1}{2 k^2}
\left(d \th^2 + \sinh^2 \th d \phi^2 \right) .  
\ee 
This is the solution of the equation, 
\be 
G_{ab} = -\Lambda g_{ab} 
\ee 
with $\Lambda = -3k^2$. The projection of the Weyl curvature on the brane
is given by [9] 
\be E_{\mu \nu} := C_{\mu\rho\nu\sigma} \, n^{\rho} \,
n^{\sigma} \ee 
where $n^{\alpha}$ is the unit normal, and we have 
\be
E_{\mu\nu} = \left( E_{tt}, E_{rr}, E_{\th\th}, E_{\phi\phi} \right) =
-\frac{k^2}{2} \left( g_{tt}, g_{rr}, - g_{\th\th}, - g_{\phi\phi}
\right).  
\ee
 Here the latin indices run from ${0,...,4}$ while the greek from ${0,...,3}$. 
Note that the brane $y=0$ is not flat in this
case and could describe a cloud of string dust of uniform energy
density [10]. Since the brane metric is not a vacuum, one might expect
from the investigation of ref.[7] that gravity may not be localized in
this case. But the metric (2) is not in the standard RS form as only the
non-angular part of the metric shares the conformal factor. This result would therefore not follow straightway. Our metric
is spatially anisotropic which is essential for the Weyl curvature to
be non-zero. If the conformal factor is shared isotropically, it would
trivially be conformally flat.

 We would now study the perturbations of the above metric for finding the bound state of massless graviton. In order to find a massless ground state mode with two degrees of freedom we will do the linear perturbative analysis of the above metric. We write the perturbed metric, $\tilde g_{ab} = g_{ab} + h_{ab}$. Since our main objective is to examine existence of the bound normalizable massless mode on the brane, we take the metric fluctuations in the extra dimension to vanish, i.e.
\be
h_{t \, y} = h_{r \, y} = h_{\th \, y} = h_{\phi \, y} = h_{y \, y} = 0 .
\ee
We further impose the following gauge conditions on the metric fluctuations on the brane
\be
\nabla^{\mu} h_{\mu \nu} = 0, \hspace{1cm} h^{\mu}{}_{\mu} = 0.
\ee
We then obtain the equation
\be
\Box h_{ab} = A^i_{ab,i} + \Gamma^i_{ij} \, A^j_{ab} - \Gamma^i_{aj} A^j_{ib} -\Gamma^i_{bj} A^j_{ia} = \Lambda\, h_{ab}
\ee
where $\Lambda $ is the cosmological constant in the bulk and 
\be
A^i_{ab} := g^{ij}\left( h_{ab,j} - \G^k_{aj} h_{kb} - \G^k_{bj} h_{ak} \right).
\ee
Here the cosmological constant in the bulk is related to $k$ by $\Lambda =
 - 3 k^2$. We look for the solutions of the form
\be
h_{a b} = e_{ab} \, \Psi(y) e^{i \, p_{j} \, x^{j}}
\ee
and further assume that the the polarization tensor $e_{ab}$ is orthogonal to a unit
timelike vector $u^a = (1,0,0,0,0)$ i.e. $e_{ab} \, u^b = 0$. It is interesting
 to note that our form of the metric clearly demarcates the physical modes of
propagation. This happens because propagation vector is null which only allows
$h_{\th\th}$ and $h_{\th\phi}$ as the physical modes, where $h^{\th}{}_{\th} =
- h^{\phi}{}_{\phi}$ due to traceless condition. It is easy to check that other modes are ruled out because the null form of $p_a$ demands that $y$ should be constant. 

The gauge conditions impose constraints on the derivatives of $h_{\th\th}$ and $h_{\th\phi}$ which are 
\be
  h_{\th \phi, \phi} +\sinh^2 \th \, h_{\th\th, \th}  + \sinh 2\th \, h_{\th\th} = 0 
\ee
and
\be
h_{\th\phi, \th} - h_{\th\th,\phi} + \coth\th \, h_{\th \phi} =  0 .
\ee
Solving the wave equation and using $m^2$ as a constant for separation of
variables it can be shown that the $y$ part comes out to be
\be
\Psi^{''}(y) - 2 \, k\, \Th(y) \Psi^{'}(y) + \left(4 \, k^2 \, + m^2 \, e^{2 \,k \, |y|} \right)  \Psi(y)  = 0
\ee
where $\Th(x)$ is the unit step function,
\begin{displaymath}
\Th(x) = \left\{ \begin{array}{ll}
+ 1 & \textrm{if $x > 0$}\\
- 1 & \textrm{if $x < 0$ .}
 \end{array} \right.
\end{displaymath}

Eq.(13) is not in the form of a wave equation, however we can always
transform it into that form through the following substitution 
\be
\Psi(y) = \chi(y) \, \psi(y), \, \hspace{1cm} \chi(y)=exp \left(k
\int_{-\infty}^{\infty} \Th(x) \, dx \right) .  
\ee 
On this substitution the first derivative term in eq.(13) vanishes and we
obtain 
\be 
\psi^{''}(y) + \left[2 \, k \, \de(y) + 3 \, k^2 + m^2 e^{2
\, k \, |y|}\right] \psi(y) = 0.  
\ee
Putting $m = 0$ we obtain the
required wave equation which would tell us about the effect of extra
dimension on the massless modes of the metric fluctuations on the
brane, 
\be 
\psi^{''}(y) + \left(2 \, k \, \de(y) + 3 \, k^2 \right)\psi(y) = 0.  
\ee 
The presence of Dirac delta function in the above
equation tells us that we should look for the solutions which are the
functions of $|y|$. The above equation has an irregular singular point
at $y=0$, hence in order to look for its solution let us consider a
general equation of the form 
\be 
f^{''}(y) + \left(a \, \de(y) + b\right) f(y) = 0 
\ee 
and look for solutions $f = f(|y|)$. Then it can
be easily seen that the solution would exist only if $a^2 = - 4 b$
with $f = exp(- a \,|y|/2)$.

The wave equation describing $y$ dependence of massless mode of the
metric fluctuation in the RS model is given by 
\be \psi^{''}(y) + \left(4 \, k \, \de(y) - 4 \, k^2 \right) \psi(y) = 0.  
\ee 
With $a = 4 \, k$ and $b = -4 \, k^2$ the RS equation satisfies the constraint
$a^2 = - 4 b$ and gives the ground state wavefunction as $exp(-2 \,k
\,|y|)$. However, in our case, eq.(16) yields $a = 2 \, k$ and $b = 3
k^2$ which implies that the constraint condition is not satisfied and
hence there can exist no solution of the form $f = f(|y|)$.

We would like to argue that a brane world model should admit all
solutions of the Einstein space equation in the bulk and not only the
specific anti - de Sitter solution. The really interesting situation
would be presented by the conformally non flat bulk solution which
would project on the brane free gravitational field as trace free
matter giving rise to ``tidal'' charge which would make non trivial
contribution. It is a pity that the most exciting aspect of the model
is not realizable in the present form of its formulation. It may be
recalled that an interesting solution in terms of the
Reissner-Nordstrom (RN) metric was obtained for a black hole on
the brane [11] precisely due to the contribution of the bulk Weyl
curvature on the brane. It turns out that the Nariai metric (2)
projects on the brane the stresses precisely of the form for which the
RN metric is the exact solution. Thus the Nariai metric (2) gives the
bulk spacetime metric corresponding to the black hole solution [11] on
the brane.

We have thus shown that the brane model described by eq.(2) does not
allow the existence of the bounded massless mode. Since that does not
happen there is no question of recovering the $1/r$ Newtonian
potential between two massive particles on the brane. That means a non
conformally flat bulk spacetime with negative $\Lambda$ does not yield
the conventional gravity in the 4-dimensions on the brane. At the same time our metric
is an exact good solution of the Einstein space equation in the
bulk. This is however hardly surprising once we notice the kind of
fine tuning which is required between $a$ and $b$ to solve an ill
behaved differential equation like eq.(17). The RS metric is just an
example of that in which the parameters just fit perfectly. Any
deviation from the specific matching values would lead to non-confinement as is clear from eq.(17). The brane world model thus seems to be very brittle for inclusion of non-zero Weyl curvature in the bulk.

By studying gravity on the branes through the $4$-dimensional
discontinuity caused by the $Z_2$ symmetric bulk, it has been
shown that if the bulk is an Einstein space near the brane,
Einstein's equation could approximately be recovered on the brane. On
the other hand if it is required to be exactly the anti - de Sitter
space, Einstein's equation cannot hold even approximately unless the
matter obeys a very contrived and artificial equation of state
[12]. And we show that if it is not anti - de Sitter, gravity cannot
remain confined to the brane. 

\vspace{0.5cm}

{\it Acknowledgment:} We thank S. Shankaranarayanan and T. Padmanabhan for 
helpful discussions as well as for reading the manuscript. PS thanks Council for Scientific \& Industrial Research for grant 
number: 2 - 34/98(ii)E.U-II.


\newpage
\begin{references}
\bibitem[a]{e-mail:} param@iucaa.ernet.in
\bibitem[b]{e-mail:} nkd@iucaa.ernet.in

\bibitem{} N. Arkani-Hamed, S. Dimopolous and G. Dvali, Phys. Lett.
{\bf B429}, 263 (1998); I. Antoniadis, N. Arkani-Hamed and
S. Dimopolous, Phys. Lett. {\bf B436}, 257 (1998); N. Arkani-Hamed and
S. Dimopolous, Phys. Rev.  {\bf D59}, 086004 (1999).
\bibitem{} G. Shiu and S. H. Henry Tye, Phys. Rev. {\bf D58}, 106007 (1998).
\bibitem{} L. Randall and R. Sundrum, Phys. Rev. Lett. {\bf 83}, 4690 (1999).
\bibitem{} C. D. Hoyle, et.al., \emph{Sub-millimeter tests of the gravitational 
inverse square law: A search for ``large'' extra dimensions}, hep-ph/0011014; 
D. j. H. Chung, et.al, \emph{Experimental Probes of the Randall-Sundrum Infinite 
Extra Dimension}, hep-ph/0010103.
\bibitem{} D. Marolf, M. Trodden, \emph{Black Holes and Instabilities
of Negative Tension Branes}, hep-th/0102135. M.G. Santos, F. Vernizzi,
P.G. Ferreira, \emph{Isotropization and instability of the brane},
hep-ph/0103112.P. Binetruy, J.M. Cline, C. Grojean, Phys.Lett. {\bf
B489} 403, (2000).
\bibitem{} T. Boehm, R. Durrer and C. van de Bruck, \emph{Dynamical
Instabilities of the Randall-Sundrum Model}, hep-th/0102144.
\bibitem{} T. Padmanabhan and S. Shankaranarayanan, \emph{Vanishing
of cosmological constant in nonfactorizable geometry}, hep-th/0011159.
\bibitem{} H. Nariai, Sci. Rep. Tohoku Univ. {\bf 34}, 160 (1950).
\bibitem{} T. Shiromizu, K. Maeda, M. Sasaki, Phys.Rev.  {\bf D62}
024012 (2000).
\bibitem{} N. Dadhich, \emph{On product spacetime with 2-sphere of
constant curvature}, gr-qc/0003026.
\bibitem{} N. Dadhich, R. Maartens, P. Papadopoulos and V. Rezania, 
Phys. Lett. {\bf B487}, 1 (2000).
\bibitem{} N. Deruelle and J. Katz, \emph{Gravity on branes}, gr-qc/01040007.
\end{references}
\end{document}